\documentclass[12pt]{article}

\usepackage{amssymb}
\usepackage{amsfonts}
\usepackage{pstricks}

\newtheorem{theorem}{Theorem}
\newtheorem{lemma}[theorem]{Lemma}

\begin{document}

\title{Quinary lattice model \\  of secondary structures of polymers}

\author{S.V.Kozyrev, I.V.Volovich}

\maketitle

\begin{center}
Steklov Mathematical Institute,

Russian Academy of Sciences,
Moscow, Russia
\end{center}

\begin{abstract}
In the standard approach to lattice proteins the models based on nearest neighbor interaction are used.
In this kind of models it is difficult to explain the existence of secondary structures --- special preferred
conformations of protein chains.

In the present paper a new lattice model of proteins is proposed which is based on non-local cooperative
interactions. In this model the energy of a conformation of a polymer is equal to the sum of energies of conformations of fragments of the polymer chain of the length five.

It is shown that this quinary lattice model is able to describe at qualitative level
secondary structures of proteins: for this model all conformations with minimal energy are combinations of lattice models of alpha--helix and beta--strand. Moreover for lattice polymers of the length not longer that 38 monomers we can describe all conformations with minimal energy.

\end{abstract}

\newpage

\section{Introduction}

Lattice models of polymers (in particular proteins) were extensively discussed in the literature, cf. \cite{GrosbergKhokhlov}.
For a review of physics of proteins see \cite{FP}.

The standard lattice models of polymers (in particular the HP model \cite{Dill2,Dill3}) describe nearest neighbor interaction of monomers (see the discussion in the next section). Conformations with low energy in the standard model have the form of compact globules. In proteins native conformations usually are combinations of secondary structures --- special preferred regular conformations, in particular alpha--helices and beta--sheets. The aim of the present paper is to introduce a lattice polymer model (quinary lattice model) where energy minima will be combinations of lattice secondary structures (models of alpha--helices and beta--strands).

The energy of a polymer in this model will be equal to the sum over conformations of fragments of a polymer of the length five. Therefore instead of consideration of explicit interactions between amino acids in a protein we consider cooperative interaction in a polymer chain. We fix a set of conformations of fragments of polymer chains and construct low energy conformations of polymers as combinations of conformations of fragments from the mentioned set.

The crucial point of our lattice model is that for two conformations of fragments of a polymer it is not always possible to construct a larger fragment which will contain both shorter fragments. This implies a set of selection rules for conformations of neighboring short fragments of a polymer. Therefore the number of possible conformations with low energy of lattice polymers in the introduced model will be small in comparison with standard models of lattice polymers with contact interaction where all compact globules will have sufficiently low energy. Actually in this paper all possible conformations with minimal energy  for sufficiently short (with the length less that 39) lattice polymers are enumerated.

Let us stress that we do not pretend that the proposed model gives realistic approximation of conformations of real proteins. This model is a mathematical toy model which describes the effect of creation of secondary structures in lattice polymers by non-local cooperative interaction.

Let us compare the approach of the present paper and some of the known results in lattice (and some off-lattice) models of proteins.

In  \cite{PandeGrosbergTanaka,ShakhGutin} design of sequences of polymers which fold to given conformations was discussed.
Review of recent results in this direction one can find in \cite{Shakhnovich2006,Baker2012}.

In \cite{KhoKha} the model of HP (hydrophobic-polar) copolymers was considered.
Primary structures of the so called protein-like copolymers (with globules which have hydrophobic core and polar shell)
were investigated, long-range correlations in the primary structures were found.

In \cite{Tang} the lattice HP-model of polymer was studied. Compact structures of 27-mer (which lie in $3\times 3\times 3$ cube)
were enumerated and number of sequences which minimize the energy in the given compact conformation was computed
for each of of the mentioned structures. It was discussed that the highly designable structures
(corresponding to large number of sequences) exhibit certain geometrical regularities which are reminiscent of
the secondary structures in natural proteins.

For other lattice and off lattice polymer models see \cite{Scheraga}, \cite{BJST}, \cite{OnuchicWolynes}, \cite{AveBikNe}.

In the mentioned above paper lattice secondary structures were discussed in the framework of the long-range order and low entropy. In particular helix like conformations in these models are combinations or right and left handed helix patterns. In the present paper we show that a special cooperative interaction can explain not only long range order but also the formation of very particular conformations, in particular, helix conformations in our approach are right--handed only.

In \cite{Nekrasov1} proteins were considered as combinations of short fragments (in particular 5--tuples) of amino acids, in \cite{Nekrasov} the statistics of conformations of fragments of proteins was discussed.

In the work \cite{Unger}, see also \cite{Micheletti}, \cite{Kolodny}, \cite{Baker}, \cite{Bujnicki} it was proposed to consider a small data
bank of (off--lattice) protein fragments that can be used as elementary building blocks to reconstruct virtually all native protein
structures.
It was found that oligomers of short length (usually between 5 and 20) found in a coarse grained representation of
native structures of proteins do not vary continuously but gather in few clusters. These clusters can be represented by structural fragments, or oligons, which can be considered as centers of these clusters. Let us note that the characteristic length of an oligon in \cite{Micheletti} is five or six amino acid residues, and in the present paper we use fragments of a lattice polymer of the length five.

The structure of the present paper is as follows.

In section 2 we introduce the quinary lattice model of a polymer and show that the lattice models of alpha--helix and beta--strand are the energy minima for the introduced model.

In section 3 we describe minima of energy of the proposed model and show that these conformations can be considered as combinations of lattice alpha--helices and beta--strands.

In section 4 we consider the heteropolymer version of the introduced model and show that in this case we can describe polymers which possess a native tertiary structure --- a conformation which minimizes the energy and is uniquely defined by the sequence of the polymer. The obtained tertiary structure is a combination of secondary structures described in sections 2 and 3.

In section 5 we discuss the model of energy of a lattice polymer which is a combination of the quinary lattice model and the standard model of the nearest (in the lattice) neighbors interaction of amino acids.

In section 6 we give a conclusion of our results.

\section{The quinary lattice model}

The standard model of energy of lattice polymers has the following form, see  \cite{GrosbergKhokhlov}. One considers a linear lattice polymer (a finite sequence of monomers connected by edges of the length one), the monomers are situated at vertices of the cubic lattice $\mathbb{Z}^3$. A conformation of a polymer of the length $N$ is a sequence of neighboring vertices without self--intersections in the cubic lattice $\mathbb{Z}^3$, i.e. the injective map
\begin{equation}\label{Gamma}
\Gamma:\{1,\dots,N\}\to\mathbb{Z}^3,
\end{equation}
where neighboring natural numbers map to neighboring (i.e. distance one) vertices of the lattice $\mathbb{Z}^3$.
In the following we will denote $\Gamma$ also the image of this map.

Energy of the conformation $\Gamma$ in the standard model is proportional to the following sum
\begin{equation}\label{E_2}
E_2(\Gamma)=-\sum_{1\le i<j\le N}\delta(d(\Gamma(i),\Gamma(j))),
\end{equation}
where $\Gamma(i)$ is the $i$-th monomer in the polymer, $d(\cdot,\cdot)$ is the distance in $\mathbb{Z}^3$, $\delta(1)=1$, $\delta(i)=0$, $i>1$. Non zero contributions to this sum come from the contacts of pairs of monomers in the conformation $\Gamma$.

The above model takes into account only the number of contacts in the conformation but does not distinguish between the conformations with the different geometry. Therefore this model does not describe formation of secondary structures (special preferred conformations of a protein), in particular, alpha--helices and beta--strands.

In the present paper we introduce the following model of a lattice polymer. The energy of a polymer in this model will be equal to the sum of contributions where each of the contributions depends on conformation of a segment of a lattice polymer of the length 5 (i.e. which contains five monomers). Therefore the energy of a polymer of the length $N$ in conformation $\Gamma$ has the form
\begin{equation}\label{E_5}
E_5(\Gamma)=-\sum_{i=3}^{N-2}\Phi(\Gamma_i).
\end{equation}
Here $\Gamma_i$ is a conformation of the $i$-th 5--tuple of monomers in a polymer, i.e.
$$
\Gamma_i=(\Gamma(i-2),\Gamma(i-1),\Gamma(i),\Gamma(i+1),\Gamma(i+2)),\qquad i=3,\dots,N-2,
$$
where $\Gamma(i)$ is the $i$-th monomer in a polymer of the length $N$, $\Phi$ is some function of conformations of 5--tuples of monomers. The function $\Phi$ is taken to be invariant with respect to lattice rotations and translations of conformations of 5--tuples.

The intersection of the two neighboring 5--tuples $\Gamma_i$ and $\Gamma_{i+1}$ contains the four monomers  $\Gamma(i-1),\Gamma(i),\Gamma(i+1),\Gamma(i+2)$.

It is easy the see that (modulo lattice translations and rotations) there exist 30 different conformations of 5--tuples without self--intersections. We choose the function $\Phi$ as follows: $\Phi(\Gamma)$ is equal to zero for all conformations except the conformations denoted 1 and 2 for which $\Phi(1)=\Phi(2)=1$, cf. Fig. 1 (by definition $\Phi$ will be also equal to one for lattice translations and rotations of conformations 1 and 2).

In Fig. 1 the edges connect neighboring monomers, thus a segment of a polymer containing five monomers will contain four edges. Conformations 1 and 2 are segments of a right handed helix, see. Fig. 2.

\begin{figure}\label{fig1}
\begin{pspicture}(2,4)
   \psline[linewidth=2pt](0,1)(3,1)(4,2)(1,2)(1,5)
   \rput(1,6){ Conformation 1}
   \psline[linewidth=2pt](8,1)(9,2)(6,2)(6,5)(5,4)
   \rput(6,6){ Conformation 2}
\end{pspicture}
\rput(0,0){ Fig. 1: Conformations 1 and 2}
\end{figure}

Proof of the following statement is straightforward.

\begin{lemma}\label{alphabeta}
{\sl
Conformation of any 5--tuple of neighboring monomers in periodic conformations of lattice polymers in Fig. 2 is either conformation 1 or conformation 2 of Fig. 1. Therefore these periodic conformations are minima of energy (\ref{E_5}).}
\end{lemma}

The left conformation in Fig. 2 may be considered as a lattice model of $\alpha$--helix, the right conformation can be considered as a model of $\beta$--strand. Therefore in the model under consideration we observe the two most important examples of secondary structures. The helix in Fig. 2 possesses the helical symmetry --- it maps into itself with respect to combination of lattice rotation and translation.

In real proteins $\beta$--strands are stabilized by hydrogen bonds between parallel chains, therefore to have a stable conformation one needs at least two parallel $\beta$--strands. Our model neglects hydrogen bonds and considers conformations which model folds of peptide chain in secondary structures as energetically profitable. Therefore this model can not pretend to give a realistic description of conformations of real proteins. Our aim is to describe qualitatively the effect of existence of secondary structures.

In the next section we will show that the described lattice $\alpha$ and $\beta$ conformations and their combinations are all possible minima of energy for the quinary lattice polymer model (\ref{E_5}).

\begin{figure}
\begin{pspicture}(2,12)
   \psline[linewidth=2pt](0,1)(3,1)(4,2)(1,2)(1,5)(0,4)(3,4)(4,5)(4,8)(1,8)(0,7)(3,7)(3,10)(4,11)(1,11)(0,10)(0,13)
   \rput(0,14){ $\alpha$-helix}
   \psline[linewidth=2pt](5,1)(8,1)(9,2)(6,2)(6,5)(9,5)(10,6)(7,6)(7,9)(10,9)(11,10)(8,10)(8,13)(11,13)
   \rput(7,14){ $\beta$-strand}
\end{pspicture}
\rput(0,0){ Fig. 2: Lattice $\alpha$-helix and $\beta$-strand}
\end{figure}

\bigskip

\noindent{\bf Remark}\quad  The 5--tuples 1 and 2 in Fig. 1 are chosen in special way. In particular these conformations violate the mirror symmetry. The recipe how to choose these conformations is the following: we would like to reproduce the lattice helix (shown in Fig. 2) as a conformation with low energy. The way to do this in the framework of the non--local model (\ref{E_5}) is to cut a lattice helix in fragments of the length 5 and make energies of the obtained fragments low. Performing this procedure we obtain the two basic fragments 1 and 2, in this sense the described model is minimal (we will get these two fragments of the length five in arbitrary model which reproduces the helix in Fig. 2). It is easy to check that the fragment length 5 is the minimal length such that combinations of fragments of this length give a reasonable set of secondary structures.

In principle one can construct more complex models of the described type taking fragments of the polymer chain with the length five or larger and taking more than two fragments.

\section{Enumeration of minimal conformations}

Let us fix the directions in the 5--tuples of monomers in Fig. 1. We choose the beginnings and the ends of 5--tuples, see Fig. 3 (where $B$ denotes the beginning and $E$ denotes the end). The obtained conformations of directed polymers we denote $\overrightarrow{1}$ and $\overrightarrow{2}$ (the left and the right conformations in Fig. 3 correspondingly). The same conformations but with the opposite directions (from the end to the beginning) we denote $\overleftarrow{1}$ and $\overleftarrow{2}$.

\begin{figure}
\begin{pspicture}(2,4)
   \rput(0,1.3){\bf B}
   \psline[linewidth=2pt](0,1)(3,1)(4,2)(1,2)(1,5)
   \rput(1,5.3){\bf E}
   \rput(8,1.5){\bf B}
   \psline[linewidth=2pt](8,1)(9,2)(6,2)(6,5)(5,4)
   \rput(5,4.5){\bf E}
\end{pspicture}
\rput(2,0){ Fig. 3: {\bf B}, {\bf E} -- beginning and end of the conformation}
\end{figure}

Conformations of lattice polymers which are minima of the energy (\ref{E_5}),
i.e. conformations of lattice polymers for which any 5--tuple of neighboring monomers has one of the conformations
$\overrightarrow{1}$, $\overrightarrow{2}$, $\overleftarrow{1}$, $\overleftarrow{2}$ are called {\it minimal}.
Note that the neighboring 5--tuples of monomers in a polymer (or 4--tuples of edges shown in Fig. 3) have the intersection containing 4 monomers (or 3 edges). We consider these conformations as models of secondary structures in proteins.

A lattice polymer in a minimal conformation $\Gamma$ generates a sequence $\Gamma_3\Gamma_4\dots\Gamma_{N-2}$ of conformations of 5--tuples of monomers (when we read the sequence of monomers in the polymer from the beginning to the end), where $\Gamma_i\in \{\overrightarrow{1}, \overrightarrow{2}, \overleftarrow{1}, \overleftarrow{2}\}$. The conformation $\Gamma$ of a polymer can be restored from the sequence $\Gamma_3\Gamma_4\dots\Gamma_{N-2}$ of conformations of 5--tuples.

We say that a minimal conformation $\Gamma_3\Gamma_4\dots\Gamma_{M-2}$ is a continuation of a minimal conformation $\Gamma_3\Gamma_4\dots\Gamma_{N-2}$, $N<M$ if the second sequence of symbols is a segment of the first sequence (i.e. the first sequence is obtained from the second by adding some symbols in the beginning and the end).

Which sequences $\Gamma_3\Gamma_4\dots\Gamma_{N-2}$ of the conformations $\overrightarrow{1}$, $\overrightarrow{2}$, $\overleftarrow{1}$, $\overleftarrow{2}$ can be generated by minimal conformations of a lattice polymer? It is not always possible to combine a couple of conformations $\overrightarrow{1}$, $\overrightarrow{2}$, $\overleftarrow{1}$, $\overleftarrow{2}$ of 5--tuples into a single conformation of a 6--tuple due to geometric restrictions, see the next statement, which describes the selection rules for the conformations of neighboring fragments.

\begin{lemma}\label{table}
{\sl
1) Possible pairs of  conformations of neighboring 5--tuples in the sequence $\Gamma_3\Gamma_4\dots\Gamma_{N-2}$ related to some minimal conformation of a lattice polymer are described by the following table
$$
\begin{array}{|c|c|c|c|c|}\hline
~ & \overrightarrow{1} & \overleftarrow{1} & \overrightarrow{2} & \overleftarrow{2}\cr\hline
\overrightarrow{1} & - & + & + & -  \cr\hline
\overleftarrow{1} & + & - & - & - \cr\hline
\overrightarrow{2} & - & - & - & + \cr\hline
\overleftarrow{2} & - & + & + & - \cr\hline
\end{array}
$$
(i.e. for any pair of symbols denoted by $+$ in the table above there exists a minimal conformation of a lattice polymer of length 6).

\medskip

2) Any conformation described by a triple of symbols from $\{\overrightarrow{1}$, $\overrightarrow{2}$, $\overleftarrow{1}$, $\overleftarrow{2}\}$ permitted by the above table corresponds to some minimal conformation of a lattice polymer of length 7 except the triples $\overrightarrow{2}\overleftarrow{2}\overrightarrow{2}$, $\overleftarrow{2}\overrightarrow{2}\overleftarrow{2}$.}
\end{lemma}

The first statement above can be checked straightforwardly. The second statement (which forbids triples of 2-s) follows from the prohibition of self--intersections.

We will show that the above statement describes all possible geometric restrictions for lattice polymers of the length shorter or equal to 38, i.e. for any sequence of conformations of 5--tuples $\Gamma_3\Gamma_4\dots\Gamma_{N-2}$, $N\le 38$, which satisfies the selection rules, there exists the corresponding minimal conformation (without self--intersections) of a lattice polymer. Therefore in order to construct minimal conformations of sufficiently short lattice polymers it is sufficient to take into account the geometric restrictions for neighboring couples and triples of 5--tuples $\Gamma_i$.

Let us consider the following periodic sequences of conformations of 5--tuples satisfying the conditions of the above statement
$$
(\overrightarrow{1}\overrightarrow{2}\overleftarrow{2}\overleftarrow{1})\dots,\qquad (\overrightarrow{1}\overleftarrow{1})\dots.
$$
Here the periods are shown in brackets (i.e. one can iterate the sequence in brackets). The corresponding conformations of lattice polymers are shown in Fig. 4. The dotted lines in Fig. 4 are central lines of $\alpha$ and $\beta$ (the definition will be given below).

The next statement describes the sequences of conformations of 5-tuples for the lattice alpha and beta structures described in the previous section.

\begin{lemma}\label{lemma2}
{\sl   Periodic sequences of conformations of 5--tuples with the periods
$$
\alpha=(\overrightarrow{1}\overrightarrow{2}\overleftarrow{2}\overleftarrow{1}),\qquad \beta= (\overrightarrow{1}\overleftarrow{1}),
$$
see Fig. 4, correspond to conformations of a lattice polymer without self--intersections.
Moreover these conformations are lattice $\alpha$--helix and $\beta$--strand (see Fig. 2) correspondingly.}
\end{lemma}

\begin{figure}
\begin{pspicture}(2,5)
   \psline[linewidth=2pt](0,1)(3,1)(4,2)(1,2)(1,5)(0,4)(3,4)(4,5)
   \psline[linecolor=red,linewidth=4pt,linestyle=dotted](2,1.5)(2,4.5)
   \rput(2,6){ Conformation $\alpha=\overrightarrow{1}\overrightarrow{2}\overleftarrow{2}\overleftarrow{1}$}
   \psline[linewidth=2pt](8,1)(5,1)(5,4)(8,4)(9,5)(6,5)
   \psline[linecolor=red,linewidth=4pt,linestyle=dotted](6.5,2.5)(7,4.5)
   \rput(8,6){ Conformation $\beta=\overrightarrow{1}\overleftarrow{1}$}
\end{pspicture}
\rput(2,0){Fig. 4: Conformations $\alpha$ and $\beta$ with central lines}
\end{figure}

The next statement describes all minimal conformations for the model (\ref{E_5}). We show that all these conformations correspond to combinations of  $\alpha$ and $\beta$ structures and for polymers not longer than 38 all such combinations are possible (correspond to conformations without  self--intersections). For longer polymers self--intersections are possible therefore only part of combinations of $\alpha$ and $\beta$ structures correspond to minimal conformations of lattice polymers.

\begin{theorem}\label{Conformations}\quad
{\sl
1) Any minimal conformation of a lattice polymer (\ref{E_5}) with the length $N>6$  has the following form:

The corresponding sequence $\Gamma_3\Gamma_4\dots\Gamma_{N-2}$ of conformations of 5--tuples can be obtained from some sequence of $\alpha$ and $\beta$ structures, $\alpha=\overrightarrow{1}\overrightarrow{2}\overleftarrow{2}\overleftarrow{1}$, $\beta=\overrightarrow{1}\overleftarrow{1}$ by elimination of a finite number of symbols $\overrightarrow{1}$, $\overrightarrow{2}$, $\overleftarrow{1}$, $\overleftarrow{2}$ in the beginning and the end of the sequence.

\medskip

2) All conformations of a lattice polymer with the length $6<N<39$ obtained as above do not contain self intersections.

There exists a sequence of conformations of 5--tuples corresponding to the conformation of a lattice polymer of the length 39 with self--intersections.}
\end{theorem}

\noindent{\it Proof}\qquad By the definition a minimal conformation of lattice polymer of the length $N$ generates a sequence  $\Gamma_3\Gamma_4\dots\Gamma_{N-2}$ of conformations of 5--tuples of monomers, $\Gamma_i\in\{\overrightarrow{1}$, $\overleftarrow{1}$, $\overrightarrow{2}$, $\overleftarrow{2}\}$.

Let us discuss how the conformations 2 (i.e. $\overleftarrow{2}$ or $\overrightarrow{2}$) can be situated in this sequence. By the selection rules we can not have more that two consecutive symbols 2 in the sequence. A single conformation 2 can be situated either at the beginning or the end of the sequence. If the polymer is longer than six monomers the minimal conformation of this polymer can not contain a subsequence $\overleftarrow{2}\overrightarrow{2}$ (since this sequence can be continued only by 2 which is forbidden).

Therefore conformations 2 can be found inside a sequence corresponding to a minimal conformation only in pairs and for polymers longer than six monomers this pair has the form $\overrightarrow{2}\overleftarrow{2}$. Any pair of this form by the selection rules should be augmented by conformations 1 (again modulo the boundaries of the sequence), and the corresponding conformation (which corresponds to a segment of a lattice polymer of the length eight) will have the form $\alpha=\overrightarrow{1}\overrightarrow{2}\overleftarrow{2}\overleftarrow{1}$.

The part of the sequence of conformations of 5--tuples of monomers which does not contain conformations 2 contains iterations of the conformation $\beta=\overrightarrow{1}\overleftarrow{1}$. This implies the first statement.

\medskip

In order to prove the second statement let us consider Fig. 10 with a lattice polymer of the length 38 in the minimal conformation  $\alpha\beta\alpha\alpha\beta\alpha\alpha\beta\alpha\alpha\beta$. One can see that any minimal conformation which is a continuation of this conformation will have self--intersections.

We have already checked that iterations of $\alpha$ or $\beta$ structures are minimal conformations without self--intersections. Let us prove that minimal conformations of sufficiently short lattice polymers containing a mixture of $\alpha$ and $\beta$ structures also do not contain self--intersections.

Let us put in correspondence to $\alpha$ and $\beta$ structures their {\it central lines} as shown in Fig. 4 by dotted lines. For the $\alpha$--structure the central line connects the centers of the opposite faces of the cube in Fig. 4 (edges of the $\alpha$--structure will be the edges of this cube). Analogously for the $\beta$--structure the central line connects the adjacent faces of the corresponding cube.

The different combinations of $\alpha$ and $\beta$ structures can be found in the figures 5, 6, 7, 8, 9. One can find that:

1) Central line of a minimal conformation of a lattice polymer is a continuous broken line;

2) For the  joint of two $\alpha$ structures or two $\beta$ structures the central lines of joint structures will be parallel, therefore central lines of $\alpha\alpha\dots$ and $\beta\beta\dots$ will be continuous straight lines;

3) For the joint of the different structures ($\alpha\beta$ or $\beta\alpha$) the central line breaks with the angle 135$^\circ$.

\begin{figure}
\begin{pspicture}(2,8)
   \psline[linewidth=2pt](0,1)(3,1)(4,2)(1,2)(1,5)(0,4)(3,4)(4,5)(4,8)(1,8)(0,7)(3,7)
   \psline[linecolor=red,linewidth=4pt,linestyle=dotted](2,1.5)(2,7.5)

   \psline[linewidth=2pt](5,1)(8,1)(9,2)(6,2)(6,5)(9,5)(10,6)(7,6)
   \psline[linecolor=red,linewidth=4pt,linestyle=dotted](7,1.5)(8,5.5)

\end{pspicture}
\rput(0,0){ Fig. 5: $\alpha\alpha$ (left) and $\beta\beta$ (right)}
\end{figure}

\begin{figure}
\begin{pspicture}(2,7)
   \psline[linewidth=2pt](0,1)(3,1)(4,2)(1,2)(1,5)(0,4)(3,4)(4,5)(4,8)(3,7)
   \psline[linecolor=red,linewidth=4pt,linestyle=dotted](2,1.5)(2,4.5)(3.5,6)
   \rput(1,9){Conformation $\alpha\beta$}
   \psline[linewidth=2pt](8,1)(5,1)(5,4)(8,4)(9,5)(6,5)(6,8)(5,7)(8,7)(9,8)
   \psline[linecolor=red,linewidth=4pt,linestyle=dotted](6.5,2.5)(7,4.5)(7,7.5)
   \rput(8,9){Conformation $\beta\alpha$}
\end{pspicture}
\rput(1,0){Fig. 6: $\alpha\beta$ and $\beta\alpha$ with central lines}
\end{figure}

It is easy to see that if the distance between central lines is larger or equal two than there are no self--intersections of the polymer.

For obtaining a self--intersection of a minimal conformation of a lattice polymer we have to put the different parts of the central line of the conformation sufficiently close. To do this we need several breaks of the central line at contacts of $\alpha$ and $\beta$ structures. The central line have to rotate for more that $180^{\circ}$ (i.e. we need more than four breaks). One can see (this can be done by enumeration of central lines of conformations) that the shortest lattice polymer with minimal conformation with self--intersection will correspond to some continuation of the conformation in Fig. 10, for example  $\beta\beta\alpha\alpha\beta\alpha\alpha\beta\alpha\alpha\beta \overrightarrow{1}$.

Therefore any minimal conformation described at the statement above for a lattice polymer with the length not larger than 38 can be realized without self--intersections.

\bigskip

\noindent{\bf Remark}\quad The above statement shows that for a lattice polymer with the length larger than six monomers the minima of energy (\ref{E_5}) have the form of combinations of lattice $\alpha$ and $\beta$ structures, and for polymers not longer than 38 all such combinations do not have self--intersections. Therefore collective interactions in models of lattice polymers are able to describe the formation of secondary structures in proteins. Arising of self--intersections for long lattice polymers is natural --- for real proteins not all arbitrary combinations of $\alpha$ and $\beta$ structures are possible.

\begin{figure}
\begin{pspicture}(2,7)
   \psline[linewidth=2pt](0,1)(3,1)(4,2)(1,2)(1,5)(0,4)(3,4)(4,5)(4,8)(3,7)(6,7)(6,4)(7,5)(7,8)
   \psline[linecolor=red,linewidth=4pt,linestyle=dotted](2,1.5)(2,4.5)(3.5,6)(6.5,6)
\end{pspicture}
\rput(1,0){Fig. 7: Conformation $\alpha\beta\alpha$ with central line}
\end{figure}

\begin{figure}
\begin{pspicture}(2,10.5)
   \psline[linewidth=2pt](0,1)(3,1)(4,2)(1,2)(1,5)(0,4)(3,4)(4,5)(4,8)(3,7)(6,7)(7,8)(7,11)(4,11)(3,10)(6,10)
   \psline[linecolor=red,linewidth=4pt,linestyle=dotted](2,1.5)(2,4.5)(3.5,6)(5,7.5)(5,10.5)
\end{pspicture}
\rput(0,0){Fig. 8: Conformation $\alpha\beta\beta\alpha$}
\end{figure}

\begin{figure}
\begin{pspicture}(2,8)
   \psline[linewidth=2pt](0,1)(3,1)(4,2)(1,2)(1,5)(0,4)(3,4)(4,5)(4,8)(3,7)(6,7)(6,4)(7,5)(7,8)(10,8)(10,5)(11,6)(8,6)(8,9)(11,9)
   \psline[linecolor=red,linewidth=4pt,linestyle=dotted](2,1.5)(2,4.5)(3.5,6)(6.5,6)(8.5,6.5)(9.5,7.5)
\end{pspicture}
\rput(0,0){Fig. 9: Conformation $\alpha\beta\alpha\beta\alpha$}
\end{figure}

\begin{figure}
\begin{pspicture}(2,12)
   \psline[linewidth=2pt](0,8)(3,8)(4,9)(1,9)(1,12)(0,11)(3,11)(4,12)(4,15)(3,14)(6,14)(6,11)(7,12)(7,15)(10,15)(9,14)(9,11)(10,12)(13,12)
   (12,11)(12,8)(9,8)(10,9)(13,9)(13,6)(12,5)(9,5)(10,6)(10,3)(9,2)(6,2)
   \psline[linewidth=2pt](7,3)(4,3)(3,2)(3,5)(4,6)(1,6)(0,5)(0,8)
   \psline[linecolor=red,linewidth=4pt,linestyle=dotted](2,8.5)(2,11.5)(3.5,13)(6.5,13)(9.5,13)(11,11.5)(11,8.5)(11,5.5)(9.5,4)(8,2.5)
   \psline[linecolor=red,linewidth=4pt,linestyle=dotted](5,2.5)(3.5,4)(2,5.5)(2,8.5)
\end{pspicture}
\rput(2,0){Fig. 10: Conformation   $\overleftarrow{1}\beta\alpha\alpha\beta\alpha\alpha\beta\alpha\alpha\beta \overrightarrow{1}$ (38 vertices)}
\end{figure}

\section{Heteropolymers and native conformations}

In the present section we consider a heteropolymer analogue of the quinary model of lattice polymer (\ref{E_5}). We consider a lattice polymer which consists of monomers of the two kinds $A$ and $B$. This polymer has the sequence of monomers $S$ and the conformation $\Gamma$. We define the energy of the polymer by the following modification of the formula (\ref{E_5}):
\begin{equation}\label{E_5g}
E_{5'}(S,\Gamma)=-\sum_{i=3}^{N-2}\left[\#_{B}(S_i)\left(\Phi_1(\Gamma_i)+\epsilon\Phi_2(\Gamma_i)\right)+
\#_{A}(S_i)\left(\Phi_2(\Gamma_i)+\epsilon\Phi_1(\Gamma_i)\right)\right].
\end{equation}
Here  $S_i$ is the sequence of monomers in the $i$-th 5--tuple of monomers in the polymer, $\Gamma_i$ is the conformation of the $i$-th 5--tuple, the function $\Phi_1$ is equal to one for the conformation 1 in Fig. 1 and to zero for all other conformations (i.e. this is a characteristic function of the conformation 1 in the space of conformations of 5--tuples of monomers)\footnote{As in section 2 we consider the conformations modulo lattice translations and rotations.}, analogously the function $\Phi_2$ is the characteristic function of the conformation 2 in Fig. 1 (thus $\Phi=\Phi_1+\Phi_2$ in formula (\ref{E_5})). The $\#_{A}(S_i)$ denotes the number of monomers of the kind $A$ in the $i$-th 5--tuple of monomers (thus $0\le \#_{A}(S_i)\le 5$), $\epsilon$ is a positive parameter, $0<\epsilon<1$. The analogous notations are used for $\#_{B}(S_i)$ -- the number of monomers of the kind $B$ in the $i$-th 5--tuple.

With the above choice of energy a lattice polymer which contains only monomers $B$ will have the minimum of energy in the conformation equal to a single $\beta$--strand, since a $\beta$--strand contains only conformations $\overrightarrow{1}$, $\overleftarrow{1}$ of 5--tuples of monomers.

Analogously a lattice polymer which consists only of monomers $A$ will maximize the number of conformations $\overrightarrow{2}$, $\overleftarrow{2}$ of 5--tuples in the conformation of the polymer. This can be achieved at lattice $\alpha$--helix (in an $\alpha$--helix a half of conformations of 5--tuples of monomers are of the type 1 and a half are of the type 2). The term $\epsilon\Phi_1(\Gamma_i)$ in expression (\ref{E_5g}) for energy makes the conformations $\overrightarrow{1}$, $\overleftarrow{1}$ more profitable energetically than a conformation which is neither of the type 1 nor of the type 2.

Therefore a lattice heteropolymer with the energy (\ref{E_5g}) for some sequences $S$ of monomers $A$ and $B$ will possess a native tertiary structure  --- a conformation which is uniquely defined by the sequence $S$ of monomers (the primary structure of the polymer), minimizes the energy (\ref{E_5g}) and consists of a combination of secondary structures (lattice $\alpha$--helices and $\beta$--strands).

\bigskip

\noindent{\bf Remark}\quad It is possible to consider more complex copolymers of $A$ and $B$. In particular, for the sequence of the kind $AA\dots ABB\dots B$ it is probable that the conformation will be a combination of $\alpha$--helix and $\beta$--strand.

We do not claim that for an arbitrary sequence $S$ of monomers $A$ and $B$ for a lattice polymer with energy (\ref{E_5g}) there exists a unique native structure or energy gap between the native conformation and other conformations. For the standard model of contact interactions random copolymers of amino acids in general do not possess energy gap between the native tertiary structures and other conformations \cite{PandeGrosbergTanaka}.

\section{Interaction of secondary structures}

Let us discuss the following generalization of the introduced in the present paper model of lattice polymers. For this generalization the expression for energy of a lattice polymer is given by the following linear combination
\begin{equation}\label{sumE5E2}
E=E_5+\lambda E_2,
\end{equation}
where $\lambda>0$ is a positive parameter, $E_2$ is given by (\ref{E_2}) and $E_5$ is given by (\ref{E_5}).

The parameter $\lambda$ should be sufficiently small in order to make energetically profitable the formation of secondary structures (at least for short segments of a polymer). The estimate for possible interval for $\lambda$ is approximately $0\le \lambda\le 4/15$, since for breaking of secondary structure in a single vertex we get energetic loss equal to four and the energetic gain for a 5--tuple of monomers will be less or equal to $15\lambda$ (i.e. we might get maximum three additional contacts for each of the five monomers).

Even for small $\lambda$ the model (\ref{sumE5E2}) allows the formation of combinations of secondary structures (which can be considered as models of tertiary structures).

Let us consider the conformation of a lattice polymer in the form of sufficiently long lattice $\alpha$--helix. If $\lambda=0$ then this conformation will be a minimum of energy (\ref{sumE5E2}).

Let us consider also the conformation which has the form of two shorter parallel $\alpha$--helices in contact (i.e. the distance between the helices is equal to one). This conformation may be considered as an $\alpha$--helix folded in half.
If $\lambda$ is larger than some threshold then the energy of a folded in half $\alpha$--helix will be larger than the energy of a single longer $\alpha$--helix since the energetic gain will contain contributions (\ref{E_2}) of contacts of a large number of monomers in the two $\alpha$--helices, and energetic loss will come from contributions (\ref{E_5}) of small number of 5--tuples of monomers in the area where $\alpha$--helix was folded in half.

Namely for the conformation with a folded in half $\alpha$--helix we will get energetic loss equal to 4 (since four 5--tuples will be not in minimal conformation) and the energetic gain equal to $(2n-1)\lambda$, where $n$ is the number of turns in each of both $\alpha$--helices (the total length $N$ of the polymer will be equal to $N=8n$). Then if $\lambda>4/(2n-1)$ the conformation with the two $\alpha$--helices in contact will be more energetically profitable than a single $\alpha$--helix. For $n\ge 4$ (i.e. the polymer length $N\ge 32$) the interval $[4/(2n-1),4/15]$ will be non empty.

We have obtained the tertiary structure of a polymer --- a compact globule built of secondary structures.

\section{Conclusion}

We construct a model of a lattice polymer (the quinary lattice model) which describes secondary structures of proteins. In this model the energy of a conformation of a polymer is equal to a sum of energies of conformations of segments of the polymer chain of the length five.

For this model with cooperative interaction all conformations with minimal energy are combinations of lattice models of alpha--helix and beta--strand. For lattice polymers of the length not longer than 38 monomers all conformations with minimal energy are described.

The introduced model of lattice proteins can be compared with the model of protein fragments considered in \cite{Unger}, \cite{Micheletti}, \cite{Kolodny}, \cite{Baker}, \cite{Bujnicki}. The crucial points of our model are the following. First, we propose the expression (\ref{E_5}) for energy of a protein for which only combinations of selected fragments will be local minima. Second, we stress the importance of the selection rules for intersections of neighboring protein fragments, which allows to reduce considerably the number of possible conformations of a lattice protein. Application of a lattice model allows to perform these investigations in explicit way.

We do not pretend that the model proposed in the present paper gives a realistic description of conformations of proteins. Our aim was to give a qualitative demonstration, at least at the level of a mathematical toy model, of importance of the mentioned selection rules for neighboring protein fragments and relation to the effect of creation of secondary structures. It is interesting that the length (equal to five) of fragments of lattice proteins used in the introduced model (this length was selected as the minimal length which makes possible the effect of formation of lattice secondary structures) coincides with the minimal length of an oligon \cite{Micheletti} in real proteins where the clustering of conformations of protein fragments was observed.

\bigskip

\noindent{\bf Acknowledgments}\qquad
This paper was partially supported by the grants of the Russian Foundation for Basic Research
RFBR 11-01-00828-a, by the grant of the President of Russian
Federation for the support of scientific schools NSh-2928.2012.1,  and
by the Program of the Department of Mathematics of the Russian
Academy of Science ''Modern problems of theoretical mathematics''.

\end{document}